\documentstyle[amsbsy,epsf,12pt]{article}

\input amssym

\newcommand{\pF}{p_{\mathrm F}}
\newcommand{\rp}{\rho_{\mathrm p}} 
\newcommand{\rn}{\rho_{\mathrm n}}

\newcounter{saveeqn}
\newcommand{\alpheqn}{\setcounter{saveeqn}{\value{equation}}\stepcounter{saveeqn}%
\setcounter{equation}{0}%
\renewcommand{\theequation}{%
              \mbox{\arabic{saveeqn}\alph{equation}}}}
\newcommand{\reseteqn}{\setcounter{equation}{\value{saveeqn}}%
\renewcommand{\theequation}{\arabic{equation}}}

\def\bild#1#2{    
        \vspace*{-5mm}
        \begin{center}
        \begin{math}
        \epsfxsize#2cm
        \epsffile{#1}
        \end{math}
        \end{center}
        }

\begin{document}

\title{Deeply bound pionic states and the effective pion mass in nuclear
systems\thanks{Work supported in part by GSI and BMBF}} 
\author{T. Waas, R. Brockmann\thanks{Permanent address: Institut f\"ur
Kernphysik, Universit\"at Mainz}, W. Weise\\ Physik-Department\\
Technische Universit\"at
M\"unchen\\ Institut f\"ur Theoretische Physik\\ D-85747 Garching, Germany}
\date{}

\maketitle

\begin{abstract}
We show that the s-wave pion-nuclear potential which reproduces the deeply
bound pionic states in Pb, recently discovered at GSI, is remarkably close to
the one constructed directly from low energy theorems based on chiral symmetry.
Converting this information into an effective pion mass we find
$m_\pi^\star/m_\pi\simeq 1.13$ in the center of the Pb nucleus, and
$m_\pi^\star/m_\pi\simeq 1.07$ in symmetric nuclear matter.
\end{abstract}

{\bf Introduction.}
The deeply bound pionic states in Pb, recently discovered at GSI \cite{GSI} and
previously predicted in refs.\ \cite{2a,2}, have stimulated renewed interest in the
nature of the s-wave part of the pion-nuclear optical potential \cite{3}. For
such states to exist with sufficiently long lifetime, there must be a subtle
cancelation between the $\pi^-$ Coulomb and strong interactions in the bulk
nucleus. In fact, the effective s-wave $\pi$-nucleon repulsion must roughly
compensate the attractive Coulomb force in the interior of the Pb
nucleus. If this were not the case, the deeply bound pion wave function would have
substantial overlap with the nuclear density distribution, and the pion would
then experience strong absorption at the nuclear surface. This
would in turn prohibit the appearance of the narrow pionic line in the
$\rm^{208}Pb(d,^3\!He)$ spectrum as it is observed in ref.\ \cite{GSI}.

The 2p and, even better, the 1s pionic bound states in such heavy systems
therefore provide strong quantitative constraints on the s-wave $\pi^-$
interactions with nucleons in nuclei, given that the p-wave interactions are
quite well established by data on pionic atoms in higher orbits \cite{3}. In
the present note we would like to update this 
discussion from the point of view of low energy theorems based on the
underlying chiral symmetry of QCD. We demonstrate that an s-wave $\pi^-$
optical potential derived from chiral symmetry alone is remarkably consistent
with the information provided by ref.\ \cite{GSI}. It is useful to
convert this information into a statement about the effective pion
mass in nuclear matter. The results confirm and sharpen our previous knowledge
about the pion-nuclear optical potential.

{\bf Pion self-energy and effective mass in matter.}
We start with a reminder of the pion self-energy in nuclear matter. Let $\rp$
and $\rn$ be the proton and neutron densities, respectively. To leading order in
those densities, the self-energy $\Pi$ (or optical potential $U$) of a pion
with energy $\omega$ and momentum $\vec{q}$ is given by:
\begin{eqnarray}\label{selfenergy}
  \Pi(\omega,\vec{q};\rho_{\rm p,n})&\equiv& 2\omega\, U(\omega,\vec{q};\rho_{\rm
  p,n})\nonumber\\
  &=&-T^{(+)}(\omega,\vec{q})(\rp+\rn)-T^{(-)}(\omega,\vec{q})(\rp-\rn)\,.
\end{eqnarray}
Here $T^{(\pm)}$ are the isospin even and odd $\pi$-nucleon (forward)
amplitudes. For a $\pi^-$,
\begin{equation}\label{amplitudes}
  T^{(\pm)}=\frac{1}{2}\left[T_{\rm\pi^-p}\pm T_{\rm\pi^-n}\right]\,.
\end{equation}
Pionic modes of excitation in nuclear matter are determined by solutions
$\omega(\vec{q})$ of the equation
\begin{equation}
  \omega^2-\vec{q}^2-m_\pi^2-\Pi(\omega,\vec{q};\rho)=0\,.
\end{equation}
The effective pion mass $m^\star_\pi(\rho)$ is defined by:
\begin{equation}\label{effmass}
  m_\pi^{\star 2}=m_\pi^2+{\mathrm Re}\, \Pi(\omega=m_\pi^\star,\vec{q}=0;\rho)\,;
\end{equation}
it is identified with the real part of the energy $\omega(\vec{q}=0)$ for a
pion at rest in matter.

{\bf Chiral s-wave potential.}
Chiral symmetry imposes strong constraints on the near-threshold behavior of
the amplitudes (\ref{amplitudes}). According to the low-energy theorem of
Tomozawa and Weinberg (TW) \cite{4}, the $\pi^-\rm N$ amplitudes at $\vec{q}=0$
behave as
\begin{equation}
  T_{\rm\pi^-p}(\omega,\vec{q}=0)=-T_{\rm\pi^-n}(\omega,\vec{q}=0)=\frac{\omega}{2f^2}+{\cal O}(\omega^2,m_\pi^2)\,,
\end{equation}
where $f$ is the pion decay constant taken in the chiral limit
($m_\pi\rightarrow 0$). Its physical value, $f_\pi=92.4$ MeV \cite{5}, differs
from $f$ by terms of order $m_\pi^2$. We use $f=86$ MeV in the following
\cite{6}.

The TW theorem states that\alpheqn
\begin{eqnarray}\label{LOresult}
  T^{(+)}(\omega,\vec{q}=0)&=&0+{\cal O}(\omega^2,m_\pi^2)\,,\\
  T^{(-)}(\omega,\vec{q}=0)&=&\frac{\omega}{2f^2}+{\cal O}(\omega^3,\ldots)\,.
\end{eqnarray}
\reseteqn
At threshold ($\omega=m_\pi$) one has to leading order:
\begin{equation}\label{LO}
  T^{(-)}_{\rm thr}=\frac{m_\pi}{2f^2}=1.86\rm\,fm\,.
\end{equation}
In chiral perturbation theory, non-leading corrections in
$T^{(-)}_{\rm thr}$ have been calculated to fourth order \cite{7}. These
calculations use the physical $f_\pi$ in the leading order term and then find
that the higher order corrections increase  $T^{(-)}_{\rm thr}=m_\pi/2f_\pi^2$
by about 15\%, so that a value close to eq.\ (\ref{LO}) results. For $T^{(+)}$
the second order corrections are altogether small 
but involve cancelations between large pieces, the $\pi\rm N$ sigma term and
other ${\cal O}(\omega^2)$ terms. In comparison with the empirical threshold
amplitudes\footnote[1]{The $T$-amplitudes are related to scattering lengths by
$T_{\rm thr}=4\pi(1+\frac{m_\pi}{M_{\rm N}})a$. The sign convention is such that
$a<0$ implies repulsion, $a>0$ corresponds to attraction.},\alpheqn
\begin{eqnarray}\label{empvalue}
  T^{(+)}_{\rm thr}&=&(-0.22\pm 0.15)\, \rm fm\,,\quad\mbox{(Sigg et
  al. \cite{8})}\nonumber\\ 
  &&(-0.16\pm 0.06)\, \rm fm\,;\quad\mbox{(KH, \cite{8a})}\\ 
  T^{(-)}_{\rm thr}&=&(1.96\pm 0.14)\, \rm fm\,,\quad\mbox{(Sigg et
  al. \cite{8})}\nonumber\\ 
  &&(1.87\pm 0.04)\, \rm fm\,;\quad\mbox{(KH, \cite{8a})} 
\end{eqnarray}
\reseteqn
the chiral leading order results (\ref{LOresult},b) are already remarkably
close, and we shall use them as reference points.

Note that at this level, with $T^{(+)}=0$, the s-wave optical potential is
simply
\begin{equation}
  U_{\rm s}=\frac{\rn-\rp}{4f^2}\simeq 44\,{\rm MeV}\,\left(\frac{\rn -\rp}
  {\rho_0} \right)\,.
\end{equation}
Here we have introduced the density of symmetric nuclear matter,
$\rho_0=0.17\mbox{ fm}^{-3}$, as a convenient scale.

Given that $T^{(+)}\simeq 0$, rescattering corrections are important as already
pointed out in ref.\ \cite{9}. For symmetric nuclear matter, the $T^{(+)}$ in
eq.\ (\ref{selfenergy}) at 
$\vec{q}=0$ is to be replaced by
\begin{equation}
  T^{(+)}_{\rm eff}(\omega,\vec{q}=0)=T^{(+)}(\omega,\vec{q}=0)-
  \left[T^{(+)^2}+2T^{(-)^2}\right]\langle
  \frac{{\mathrm e}^{{\mathrm i}kr}}{4\pi r}\rangle\,, 
\end{equation}
where the averaged spherical pion wave with $k=\sqrt{\omega^2-m_\pi^2}$
propagates between pairs of nucleons. For a Fermi gas we have
$\langle{\mathrm e}^{{\mathrm i}kr}/r\rangle\simeq 3\pF/2\pi+{\mathrm i}k$, where
$\pF=(3\pi^2\rho/2)^{1/3}$ is the Fermi momentum, and we can drop the
small ${\mathrm i}k$ term for $k\ll \pF$. Using eqs.\ (6) in leading
chiral order we find
\begin{equation}\label{teff}
  T_{\rm eff}^{(+)}(\omega,\vec{q}=0)=-3\pF\left(\frac{\omega}{4\pi
  f^2}\right)^2 
  \simeq-0.36\frac{\omega^2}{m_\pi^2}\left(\frac{\rho}{\rho_0}\right)^{1/3}
  \,\mbox{fm}\,.\nonumber
\end{equation}
For asymmetric nuclear matter we can still assume approximately equal inverse
correlation lengths $\langle 1/r\rangle$ for protons and neutrons and express
them in terms of a common Fermi momentum determined by
$\rho=\rp+\rn$. Additional rescattering corrections proportional to
$T^{(+)}T^{(-)}\langle 1/4\pi r\rangle (\rn-\rp)$ vanish for $T^{(+)}=0$ and
can safely be ignored when using  $T^{(+)}$ from eq.\ (\ref{empvalue}).
Then the self-energy $\Pi=2\omega\, U=-T^{(+)}_{\rm
eff}(\rp+\rn)-T^{(-)}(\rp-\rn)$, taken at $\omega=m_\pi$ and $\vec{q}=0$, gives
the threshold 
s-wave optical potential
\begin{equation}\label{chiral}
  U^{(0)}_{\rm s}\simeq 8.5\mbox{ MeV}\,\left(\frac{\rp+\rn}{\rho_0}\right)^{4/3}+
  44\mbox{ MeV}\,\left(\frac{\rn-\rp}{\rho_0}\right)\,. 
\end{equation}
We refer to eq.\ (\ref{chiral}) in the following as the ``chiral'' s-wave potential. A
non-zero $T^{(+)}_{\rm thr}\simeq-0.1$ fm would add a correction $\Delta
U_{\rm s}\simeq 2\mbox{ MeV}\, 
(\rp+\rn)/\rho_0$ to eq.~(\ref{chiral}).

{\bf Effective pion mass.}
The effective pion mass $m^\star_\pi(\rho_{\rm p,n})$ resulting from eq.\
(\ref{effmass}) when one uses the ``chiral'' s-wave potential (\ref{chiral}),
are shown in 
Fig.\ 1 for symmetric matter and for a typical example of asymmetric
matter. Clearly the mass shift is small for matter with $\rm N=Z$, less than 10\% of
the free mass at $\rho=\rho_0$. For systems with a large neutron excess the
effect of $T^{(-)}$ takes over, and this is obviously relevant for nuclei such
as Pb.

The detailed analysis of $\pi$-nuclear bound states requires of course to go
beyond just the ``chiral'' s-wave potential. Apart from the Coulomb potential,
the p-wave term of the optical potential has to be added, and an absorptive
potential must be included. This is done in the
standard and time-honored way \cite{3,9}. The Klein-Gordon equation to be solved in
$\vec{r}$-space is:
\begin{equation}
  \left[\vec\nabla^2-m_\pi^2+\left(\omega-V_{\rm c}(\vec{r})\right)^2-2\omega\, U(\omega,\vec{r})\right]\phi(\vec{r})=0\,,
\end{equation}
where $V_{\rm c}(\vec{r})$ is the Coulomb potential generated by the charge
distribution $\rho_{\rm p}(\vec{r})$. For the optical potential $U=U_{\rm
s}+U_{\rm p}$ we use the chiral s-wave potential as before but add a
phenomenological absorption term: 
\begin{equation}\label{s-wave}
  U_{\rm s}(\vec{r})=U_{\rm s}^{(0)}(\vec{r})+B\rho^2(\vec{r})\,,
\end{equation}
where $U_{\rm s}^{(0)}(\vec{r})$ is given by eq.\ (\ref{chiral}) but now with
local density distributions $\rho_{\rm p,n}(\vec{r})$. The $B\rho^2$ term is
introduced as usual to parameterize absorption effects, with ${\mathrm Im}\,
B=-0.27\,m_\pi^{-5}$ fitted to the widths of a large amount of pionic atom
levels in higher orbits. We choose ${\mathrm Re}\, B=0$ in our standard
set. The nonlocal p-wave term is of the form 
\begin{equation}\label{p-wave}
  U_{\rm p}(\vec{r})=\frac{2\pi}{m_\pi}\vec{\nabla}F(\vec{r})\vec{\nabla},
\end{equation}
with canonical input for the complex function $F(\vec{r})$ as specified in
ref.\ \cite{3}.

In table 1 we show results for the energies and widths of deeply bound 1s and 2p
states of the $(\pi^-\,^{207}{\rm Pb})$ system as calculated with the optical
potential (\ref{s-wave}, \ref{p-wave}). The proton and neutron density
distributions have been obtained from a realistic Skyrme-Hartree-Fock
calculation \cite{10} which reproduces the measured charge distribution of
$^{208}{\rm Pb}$.

Evidently, the chiral s-wave potential (\ref{chiral}) works remarkably well
when combined with the 
non-local p-wave potential (\ref{p-wave}) and the absorptive parts. In table 1
we have also examined the sensitivity to changes of 
$T^{(+)}_{\rm eff}$ by adding a correction $\delta T_{\rm eff}^{(+)}$ to eq.\
(\ref{teff}). The best fit to the deeply bound states in Pb is found for values
of $\delta T_{\rm eff}^{(+)}$ well within the empirical range of uncertainties in
eq.\ (\ref{empvalue}). There seems to be no need for a substantial dispersive
real part, ${\mathrm Re}\, B$, in the s-wave absorptive potential.
The ``chiral'' effective s-wave potential for $^{207}{\rm Pb}$  
produces deeply bound $\pi^-$ states with $E_{\rm 2p}=-5.39$ MeV and $E_{\rm
1s}=-7.27$ MeV. The widths of these states are smaller than 1 MeV.

We have checked the stability of these results with respect to changes of the
p-wave potential and the s-wave absorptive part by using several
parameterizations available in the literature \cite{11}. The differences are
marginal as long as these potentials fit the large amount of pionic atom data
for higher orbits.

The effective pion mass profile resulting from the self-consistent solution of
\begin{equation}\label{profile}
  m_\pi^{\star 2}(r)=m_\pi^2+2\omega {\mathrm Re}\, U_{\rm s}(r)
\end{equation}
for Pb in the absence of the Coulomb potential is shown in Fig.\ 2. In the
nuclear center the effective mass increases
by a little less than 20 MeV as compared to its free space value. About half of
this shift comes from $T^{(+)}_{\rm eff}$, the other half results from
$T^{(-)}$ and the neutron excess.

We conclude that most of the weakly repulsive s-wave $\pi^-$-nuclear potential
can indeed be directly understood in terms of basic theorems derived from
chiral symmetry. The new data on deeply bound pionic states have sharpened the
quantitative constraints on this potential considerably. In particular, the
widths of these states are a sensitive measure of the subtle balance between
s-wave repulsion and Coulomb attraction.

{\bf Acknowledgments}

We would like to thank P. Kienle, T. Yamazaki and H. Gilg for fruitful
discussions on the data and their interpretation.

\newpage
 
\begin{center}
\begin{tabular}{r@{}l|cc|cc}
\multicolumn{2}{c|}{\rule[-2mm]{0mm}{6mm} $\delta T_{\rm eff}^{(+)}\,[{\rm
fm}]$}&$E_{\rm 1s}$ [MeV]&$\Gamma_{\rm 1s}$ [keV]&$E_{\rm 2p}$
[MeV]&$\Gamma_{\rm 2p}$ [keV]\\ \hline
-0&.10&-7.135&787&-5.313&567\\
-0&.05&-7.200&871&-5.352&630\\
0&&-7.266&968&-5.391&702\\
+0&.05&-7.333&1077&-5.430&784\\
+0&.10&-7.400&1198&-5.468&876\\ \hline
\multicolumn{2}{c|}{exp.\ \cite{GSI}}& & &$-5.4\pm 0.2$& $<0.8$ MeV \\ \hline
\end{tabular}
\end{center}
{\bf Table 1:} Energies $E=\omega -m_\pi$ and widths $\Gamma$ for pionic 1s and
2p states in $^{207}$Pb calculated with the optical potential (\ref{s-wave},
\ref{p-wave}). The ``standard'' set with $\delta T_{\rm eff}^{(+)}=0$ uses the
chiral s-wave potential $U_{\rm s}^{(0)}$ of eq.\ (\ref{chiral}) and ${\mathrm
Re}\, B=0$ Results obtained with a correction $\delta T_{\rm eff}^{(+)}$ to
$T_{\rm eff}^{(+)}$ of eq.\ (\ref{teff}) are also shown. The experimental data
from ref.\ \cite{GSI} are displayed at the bottom of the table.

\bigskip

\bigskip

\bigskip

{\bf Figure Captions:}

Fig.\ 1: Effective pion mass in symmetric nuclear matter ($x=1$, dashed line)
and an example of asymmetric matter ($x=\rn/\rp=1.6$, solid line) as a function
of density $\rho=\rp+\rn$, calculated with the ``chiral'' s-wave potential
(\ref{chiral}).

Fig.\ 2: Profile of the effective $\pi^-$ mass in $^{207}$Pb deduced from eq.\ 
(\ref{profile}).

\newpage

\bild{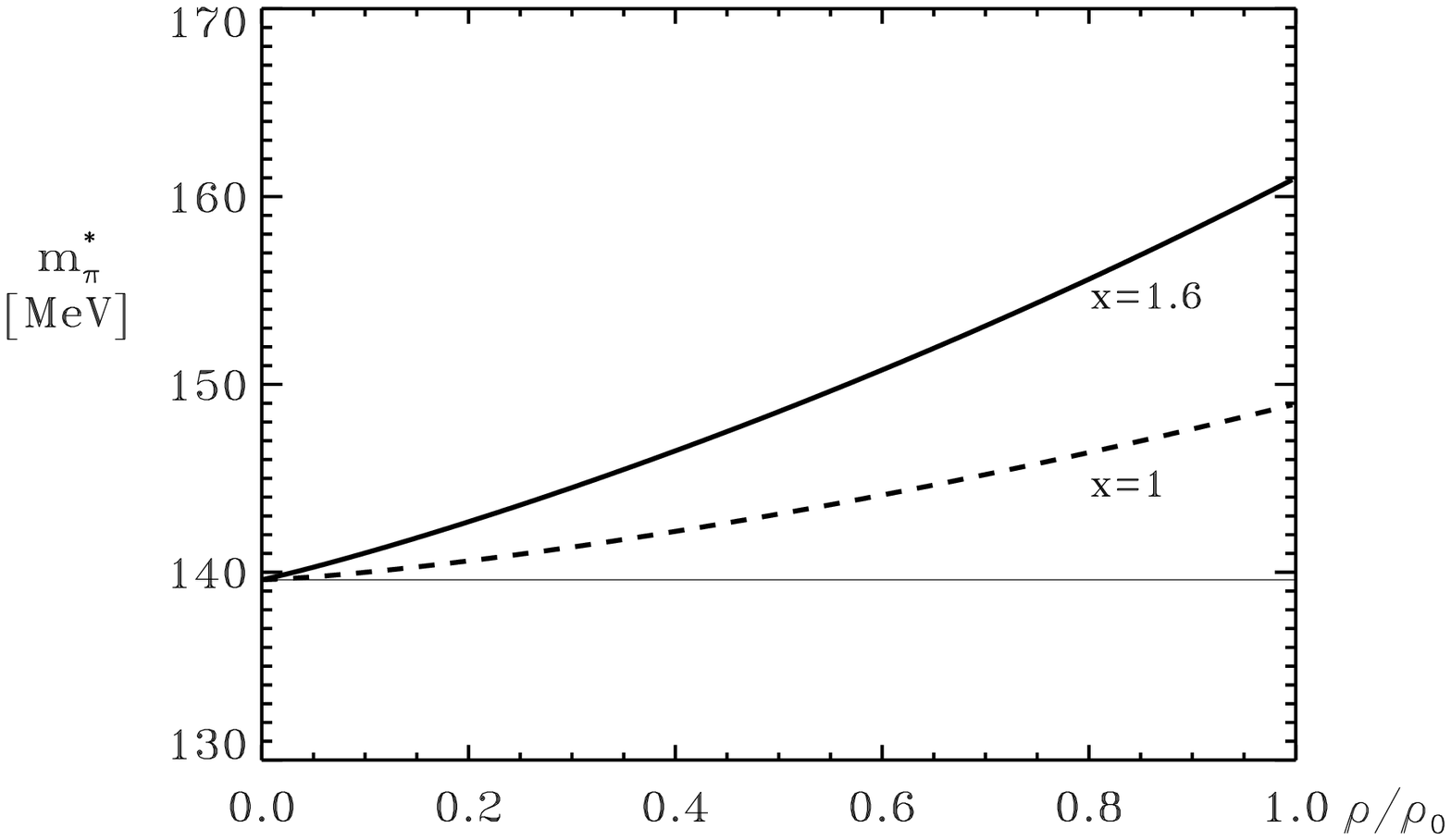}{14}
\vspace*{-.5cm}
\centerline{\bf Figure 1}

\bigskip

\bild{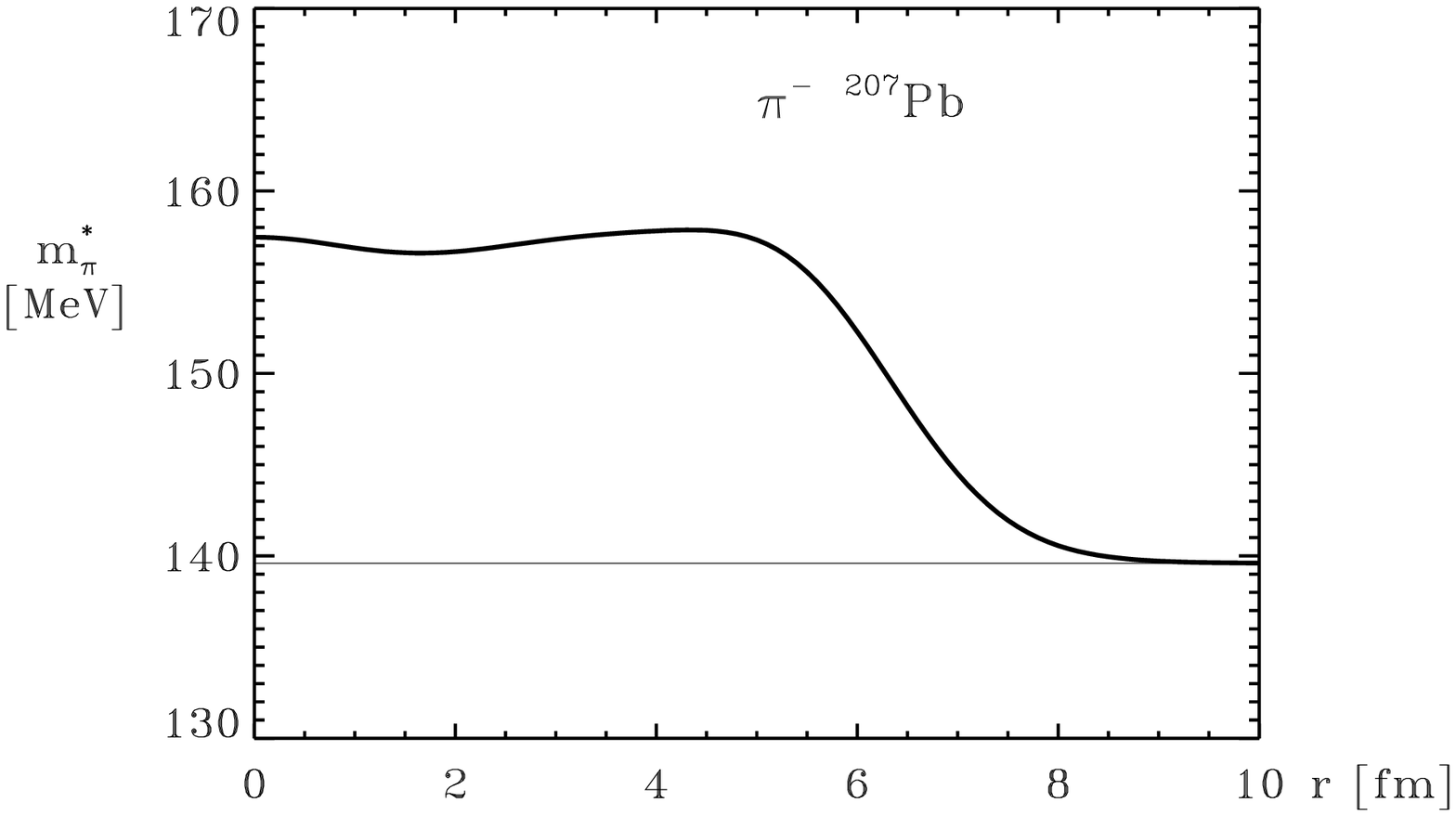}{14}
\vspace*{-.5cm}
\centerline{\bf Figure 2}

\begin{thebibliography}{99}
\bibitem{GSI} T. Yamazaki et al., {\em Z. Phys.\/} {\bf A355} (1996) 219
\bibitem{2a} E. Friedman and G. Soff, {\em J. Phys.\/} {\bf G11} (1985) L37
\bibitem{2} H. Toki and T. Yamazaki, {\em Phys.\ Lett.\/} {\bf B213} (1988)
129;\\H. Toki, S. Hirenzaki, T. Yamazaki and R.S.\ Hayano, {\em Nucl.\ Phys.\/}
{\bf A501} (1989) 653
\bibitem{3}  T. Ericson and W. Weise, {\em Pions and Nuclei\/} (Clarendon
Press, Oxford, 1988) 
\bibitem{4} Y. Tomozawa, {\em Nuovo Cim.\/} {\bf 46A} (1966) 707;\\
 S.Weinberg, {\em Phys.\ Rev.\ Lett.\/} {\bf 17} (1966) 616 

\bibitem{5} Review of Particle Physics, {\em Phys.~Rev.\/} {\bf D54} (1996) 319
\bibitem{6} J. Gasser and H. Leutwyler, {\em Ann.\ Phys.\/} {\bf 158} (1984) 142

\bibitem{7}V. Bernard, N. Kaiser and Ulf-G.~Mei{\ss}ner, {\em Phys.~Lett.\/} {\bf
B309} (1993) 421;\\ V. Bernard, N. Kaiser and Ulf-G.~Mei{\ss}ner, {\em
Phys.~Rev.\/} {\bf C52} (1995) 2185;
\bibitem{8} D. Sigg et al., {\em Phys.\ Rev.\ Lett.\/} {\bf 75} (1995) 3245, {\em Nucl.\
Phys.\/} {\bf A609} (1996) 269
\bibitem{8a} R. Koch, {\em Nucl.\ Phys.\/} {\bf A448} (1986) 707;
\bibitem{9} M. Ericson and T. Ericson, {\em Ann.\ Phys.\/} {\bf 36}
  (1966) 323
\bibitem{10} K. Pomorski, private communication
\bibitem{11} L. Tauscher, in: {\em Physics of exotic atoms}, Erice 1977 (ed.\
G. Fiorentini and G. Torelli), p.\ 145;\\
C.J.\ Batty et al., {\em Phys.\ Rev.\ Lett.\/} {\bf 40} (1978)
931;\\ C.J.\ Batty, E. Friedman and A. Gal, {\em Nucl.\ Phys.\/} {\bf A402}
(1983) 411;\\ J. Carr, H. McManus and K. Stricker, {\em Phys.\ Rev.\/} {\bf
C36} (1987) 681 
\end{thebibliography}
\end{document}